# Regression Testing of Virtual Prototypes Using Symbolic Execution

**Bin Lin[1], and Dejun Qian[2]**

[1] Computer Science Department, Portland State University
Portland, OR 97207, USA
linbin@cs.pdx.edu

[2] Intel Corporation
Hillsboro, OR 97124, USA
dejun.qian@intel.com

**Abstract**

Recently virtual platforms and virtual prototyping techniques have been widely applied for accelerating software development in electronics companies. It has been proved that these techniques can greatly shorten time-to-market and improve product quality. One challenge is how to test and validate a virtual prototype. In this paper, we present how to conduct regression testing of virtual prototypes in different versions using symbolic execution. Suppose we have old and new versions of a virtual prototype, we first apply symbolic execution to the new version and collect all path constraints. Then the collected path constraints are used for guiding the symbolic execution of the old version. For each path explored, we compare the device states between these two versions to check if they behave the same. We have applied this approach to a widely-used virtual prototype and detected numerous differences. The experimental results show that our approach is useful and efficient.

## 1. Introduction

Nowadays there has been great pressure on electronics product developers to shorten the time-to-market and improve the product quality. However, time-to-market and product quality are usually opposing attributes of a product development process. Developers can shorten the time-to-market by skipping validation steps, thus it can harm the product quality. On the other hand, if developers want to improve product quality, more validation effort need to be devoted which means more time required. This demands innovative approaches and efficient methodologies to accelerate product development and validation to save time and improve quality.

Recently virtualization and virtual prototyping techniques have been widely used in electronics companies, such as Intel and ARM, to accelerate the product development cycle [1, 2]. These techniques provide a convenient way for developers to start software development without silicon prototypes [3]. As shown in Figure 1, in the traditional product development process, software development has largely waited until the first RTL design or FPGA prototype become available. Since virtual prototypes (VP) are high-level functional models, the VP development requires less effort and can be delivered to software developers much earlier. With the virtual prototypes and virtual platform, software developers can start driver and firmware development much earlier than before. In this way, product development team can shift-left product development process and reduce the overall time. Moreover, virtual prototypes can be used as reference models for generating post-silicon functional test cases [4] and checking the correctness of physical devices [5]. Furthermore, virtual prototypes can be mixed with emulation and FPGA platform for early system integration and hardware/software validation [6, 7].

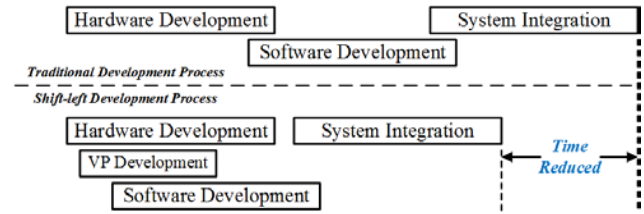

Fig 1. Traditional development VS VP-based shift-left development

Since virtual prototypes are software models developed according to the hardware specifications by developers, it is very important to validate virtual prototypes. Virtual prototypes change frequently due to feature updates, specification changes and bug fixes. To validate the changes, developers need to perform regression testing on virtual prototypes to ensure that the changes have not introduced new faults. However, applying traditional regression testing to validating virtual prototypes is difficult. This demands a new approach to regression testing of virtual prototypes.

In this paper, we propose a new regression testing approach for checking conformance between two versions of virtual prototypes. Our approach takes the new version of virtual prototype as the reference model and executes it symbolically to collect all possible path information. For each path explored, the old version of virtual prototype is executed following the path constraints collected. Then the final states are compared between the old and new versions to check if both versions conform. We have applied our approach to a widely-used virtual prototype and detected several inconsistencies in three versions of the target virtual prototype. The experimental results show that our approach is useful and efficient.

Our research makes three main contributions as follows:

    1) **Build a regression testing framework for checking two versions of a virtual prototype.** A framework is proposed for regression testing of virtual prototypes. The framework takes two different versions of a virtual prototype as inputs. Then symbolic execution is conducted and final device states are collected to detect differences between two versions.

    2) **Generate a test harness for guiding symbolic execution and checking differences.** In order to guide symbolic execution, we need to create a test harness to connect two versions of virtual prototypes together. Moreover, the developers can decide what device states they want to check in the test harness.

    3) **Evaluate on a widely-used virtual prototype.** We have evaluated our approach on a QEMU E1000 virtual network device. The experimental results show that our approach can efficiently detect differences between two different versions.

The remainder of this paper is structured as follows. Section 2 compares differences between a silicon device and the corresponding virtual prototype. Section 3 presents our approach. Section 4 demonstrates the experimental results. Section 5 discusses the related work. Section 6 concludes and discusses future work.

## 2. Silicon Device VS Virtual Prototype

A virtual prototype is a software model which emulates necessary behaviors defined in the hardware specification.

A virtual prototype should behave the same as the corresponding silicon device from the view of software developers. In order to better introduce what a virtual prototype is, we compare the differences between a silicon device and the corresponding virtual prototype. In the following, we use PCI devices as examples since PCI devices are complex and widely used.

### 2.1 Silicon Device

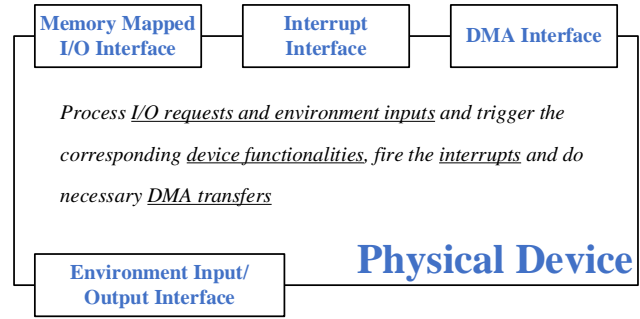

Fig. 2 An overview of a physical device

As shown in Figure 2, a physical device includes two parts: interfaces and internal functionalities. A PCI physical device commonly includes four interfaces:

1) **Memory Mapped I/O Interface**: the CPU performs write/read register operations on the device.
2) **Interrupt Interface**: the device sends electronic signals to the CPU notifying a hardware event.
3) **DMA Interface**: the device accesses the main system memory independently of the CPU.
4) **Environment Input/Output Interface**: the device sends the output data to the environment and the environment sends the input data to the device.

Inside a silicon device, all device functionalities are implemented as electronic logic to process I/O requests and environment inputs and trigger the corresponding device functionalities, fire interrupts and do necessary DMA transfers.

A silicon device is connected to the system board through the system bus. The silicon device is also connected to outside environment through different connections, such as network cables and VGA connectors.

### 2.2 Virtual Prototypes

As shown in Figure 3, a virtual prototype also includes two parts: interface functions and internal behavioral functions. A PCI virtual prototype includes four kinds of interface functions:

1) **Memory Mapped I/O Function**: when there is a register write/read CPU operation, the

corresponding interface functions are called to process the requests.

2) **Interrupt Function**: when a virtual prototype needs to fire an interrupt, the interrupt function is called to notify the virtual platform.
3) **DMA Function**: the device accesses the main system memory through DMA interface functions.
4) **Environment Input/Output Function**: the output function is used by the virtual prototype to send data while the input function is invoked by the virtual platform to notify the virtual prototype there is data received.

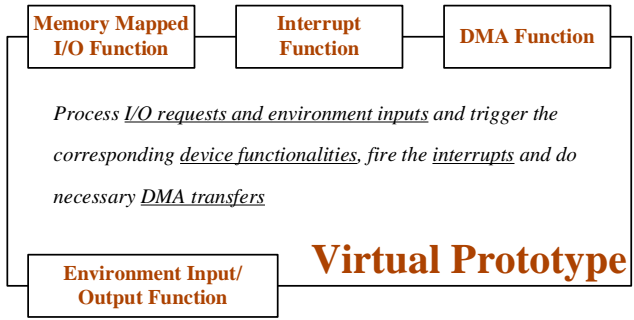

Fig 3. An overview of a virtual prototype

Inside a virtual prototype, all device functionalities are implemented as software functions to process I/O requests and environment inputs and trigger the corresponding device functionalities, fire interrupts and do necessary DMA transfers.

A virtual prototype is one component of a virtual platform. The memory mapped I/O functions and environment input functions are defined in the virtual prototype and invoked by the virtual platform to perform register access and process received data. The interrupt functions, DMA functions and environment output functions are defined in the virtual platform and called in the virtual prototype to fire interrupts, perform DMA access and send the data to the environment.

## 3. Our Approach

### 3.1 Overview

Virtual prototypes change frequently due to feature updates, specification changes and bug fixes. To validate the changes, developers need to perform regression testing on virtual prototypes to ensure that the changes have not introduced new faults. The basic idea of our approach is to detect differences between two versions of a virtual prototype. In this way, it gives developers a better understanding what features have been implemented and what bugs have been fixed in the newer version.

Suppose we have two versions of a virtual prototype $V_{old}$ and $V_{new}$, our approach can efficiently detect differences between $V_{old}$ and $V_{new}$. The basic workflow of our approach is shown in Figure 4.

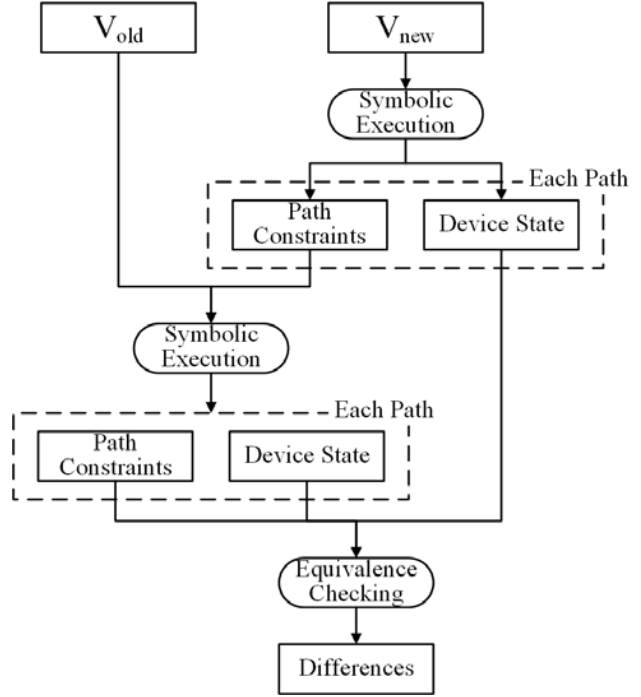

Fig 4. The workflow of our approach

Our approach takes two versions of a virtual prototypes as inputs. First, the new version of the virtual prototype is executed symbolically and all paths are explored. For each path explored, the path constraints $C$ and the final device state $S$ are collected. Guided by the collected path constraints C, the old version of the virtual prototype is executed symbolically and the path constraints $C'$ and the final device state $S'$ for each path are collected. The last step of our approach is equivalence checking. With the path constraints $C'$, we compare $S$ and $S'$ to detect the differences between $V_{old}$ and $V_{new}$.

### 3.2 Test Harness Generation

A virtual prototype is a software model which is not a standalone program. To apply our approach to regression testing of virtual prototypes, a test harness is needed to compose a complete program. The test harness mainly includes two parts:

1) To invoke the interface functions of a virtual prototype correctly, we need to construct a device state variable and device request variables. Moreover, interface functions should be called to trigger device functionalities under a desired device state upon a device request.
2) To conduct symbolic execution and equivalence checking, we need to make necessary variables as symbolic variables and invoke some special functions to guide execution and collect runtime constraints and device states.

In our case study, we use QEMU E1000 virtual prototype [8, 9]. An excerpt of the test harness we generated is shown in Figure 5.

```
int main()
{
   // Define necessary variables
   E1000State state1, state2;
   uint64_t offset, value;

   // Make symbolic states and variables
   make_symbolic_states(&state1, &state2, sizeof(E1000state));
   make_symbolic(&offset, sizeof(offset), "offset");
   make_symbolic(&value, sizeof(value), "value");

   // Invoke register write interface function
   mmio_write_new(&state1, offset, value);
   // Save the device state S
   save_state1(&state1);

   // Invoke register write interface function
   mmio_write_old(&state2, offset, value);
   // Save the device state S'
   save_state2(&state2);

   // Compare the states
   compare_states();

   return 0;
}
```

Fig 5. Excerpt of the E1000 test harness

This excerpt is not the same as the real test harness, however it shows the high-level structure. In this excerpt, only register write interface functions are invoked. In fact, we have created a complete test harness to invoke all interface functions like register read, register write and data receive interface functions. Furthermore, our approach checks not only device states but also return value after register read interface functions are invoked.

3.3 Symbolic Execution

In the paper [10], they have demonstrated how to conduct symbolic execution of virtual prototypes. Inspired by the idea, our approach also employs KLEE [11] as our symbolic execution engine and have further modified KLEE for our specific regression testing approach.

We mainly modified KLEE in four parts:
1) To construct device states for both versions of a virtual prototype and assign the same symbolic value to both states, we have added a special function "*make_symbolic_states*" and implement the corresponding handlers.
2) We have added implementations to save device states and register read return values after invoking interface states.
3) We have added the ability to compare two device states. After comparing device states, a final report is provided for further analysis.
4) We have removed all unnecessary loops in the models. Since the virtual prototype we conducted the case study on is a network device, there are some loops in the packet transmit and receive functionalities. Those loops don't affect the device state change, therefore we removed all those loops to avoid path explosion in symbolic execution.

## 4. A Case Study

4.1 Overview

We have applied our approach to a QEMU E1000 virtual network prototype. All experiments have been conducted on a physical machine with 2.5GHz CPU and 4GB memory.

In our case study, we verified if two versions of the E1000 virtual prototype have the same final states after processing external requests. Such external requests include MMIO mapped register read and write, network packet receive request. In our test harness, the corresponding interface functions are invoked separately to trigger different device functionalities and conduct difference checking.

To apply our approach, we have selected three different versions of QEMU release: 0.13.0, 1.3.1 and 2.4.1. We have also summarized some details about these three versions like lines of code (LOC) and release dates shown in Table 1. These three versions were released on different years and have different sizes.

Table 1. Summary of three different versions

| Version | LOC | Release Date |
|---------|------|--------------|
| 0.13.0  | 969  | Nov 29, 2010 |
| 1.3.1   | 1105 | Jan 28, 2013 |
| 2.4.1   | 1377 | Nov 03, 2015 |

4.2 Detected Differences

After applying our approach to three different versions, most possible paths were explored and many differences were detected. We list all results in Table 2.

Table 2. Summary of detected differences

| Old Version | New Version | # of Paths | # of Differences |
|-------------|-------------|------------|------------------|
| 0.13.0      | 1.3.1       | 148        | 7                |
| 0.13.0      | 2.4.1       | 422        | 15               |
| 1.3.1       | 2.4.1       | 428        | 13               |

As shown in Table 2, hundreds of paths have been explored and many differences have been detected. In our approach, we summarize the same differences as one unique difference. For example, if the same state differences are detected in different paths, we consider them as the same difference. Here we only show the number of unique differences.

4.3 Performance Evaluation

We also evaluate the performance of our approach. Table 3 shows the memory and time usage of our experiments.

Table 3. Summary of memory and time usage

| Old Version | New Version | Memory (MB) | Time (Min) |
|-------------|-------------|-------------|------------|
| 0.13.0      | 1.3.1       | 320         | 3          |
| 0.13.0      | 2.4.1       | 360         | 17         |
| 1.3.1       | 2.4.1       | 380         | 18         |

As shown in table 3, our experiments can finish all checking between three versions in one hour and the peak memory usage is less than 400 MB.

## 5. Related Work

Virtual platforms and virtual prototypes have been more and more utilized by different electronic vendors and academic research groups. There are many open-source and commercial virtual platforms like Simics [12] and QEMU [9] developed in the past decade [13]. Different virtual platform solutions have been developed by three large electronic design automation companies, Synopsys, Cadence and Mentor Graphics [14 - 16]. Under different virtual platforms, many virtual prototypes have been developed and utilized. These virtual prototypes cover different kinds of hardware devices, such as network, USB, audio and video.

Symbolic execution and concolic execution [11, 17 - 21] have been widely used for validating software programs. Many symbolic execution tools and related techniques have been developed and used for validating software programs and detecting security issues [22, 23]. In the past several years, symbolic execution has been further applied to hardware domain [24 - 26]. Symbolic execution of RTL design can be used for validating RTL designs [27, 28] and generating high-quality test vectors for design testing [29 - 31]. Symbolic execution of virtual prototypes has been deeply explored and utilized for coverage analysis [32], test generation [4] and conformance checking for post-silicon functional validation [5, 33, 34].

## 6. Conclusions

In this paper, we present how to apply symbolic execution to different versions of a virtual prototype for regression testing. We have detected many differences between three versions of the QEMU E1000 virtual prototype. The experimental results show that our approach can efficiently capture differences between different versions to avoid introducing faults into newer versions of the virtual prototype. In the future, we will apply our approach to more virtual prototypes.